\documentclass{llncs}

\usepackage{lmodern}
\usepackage[T1]{fontenc}

\usepackage{algpseudocode}
\usepackage{graphicx}
\usepackage[labelfont=bf,labelseparator=period,aboveskip=10pt,belowskip=-2pt]{caption}
\usepackage[labelfont=bf]{subfig}

\algrenewcommand\algorithmicindent{1.115em}
\algdef{SE}[PROC]{PROC}{CORP}{}{}
\algnotext[PROC]{CORP}
\algnotext[IF]{EndIf}
\algnotext[ELSE]{EndIf}
\algnotext[WHILE]{EndWhile}

\newenvironment{notation}%
  {\trivlist\item[\hskip \labelsep{\bfseries Notation}]\itshape}%
  {\endtrivlist}

\newcommand{\IF}{\textbf{\sffamily if}}
\newcommand{\THEN}{\textbf{\sffamily then}}
\newcommand{\ELSE}{\textbf{\sffamily else}}

\newcommand{\RETURN}{\textbf{\sffamily return}}

\title{Distributed Work Stealing for \\ Constraint Solving}

\author{Vasco Pedro \and Salvador Abreu}

\institute{Departamento de Inform\'atica,
  Universidade de \'Evora and
  \\[3pt]
  CENTRIA FCT/UNL, Portugal
  \\[2pt]
  \email{\{vp,spa\}@di.uevora.pt}}

\begin{document}

\maketitle

\begin{abstract}
  With the dissemination of affordable parallel and distributed
  hardware, parallel and distributed constraint solving has lately
  been the focus of some attention. To effectually apply the power of
  distributed computational systems, there must be an effective
  sharing of the work involved in the search for a solution to a
  Constraint Satisfaction Problem (CSP) between all the participating
  agents, and it must happen dynamically, since it is hard to predict
  the effort associated with the exploration of some part of the
  search space. We describe and provide an initial experimental
  assessment of an implementation of a work stealing-based approach to
  distributed CSP solving.
\end{abstract}

\section{Introduction}

Constraints are used to model problems with no known polynomial
algorithm, but for which search techniques developed within the field
of constraint programming provide viable procedures. Besides classical
applications, such as planning and scheduling, constraints have
recently been successfully applied in the contexts of bioinformatics
\cite{HCP:bioinf} and computer network monitoring \cite{pds@inap2009}.

Notwithstanding their relative efficiency, constraint solving methods
are computationally demanding and good candidates to benefit from
multiprocessing. Moreover, the declarative style of constraint
programming frees the programmer from concerns usually entailed by
parallel and distributed programming, such as control,
synchronisation, and communication issues. In fact, the programmer may
not even be aware that there is any parallelism involved in solving
the problem.

Given the increasing availability of parallel computational resources,
in the form of multiprocessors, clusters of computers, or both, there
is a need for an effective way to help incorporating that power into
the constraint programming setting. In this context, our goal is to
build a library which takes advantage of parallel hardware in a
transparent way, for constraint solving.

In parallel constraint solving
\cite{lsf:tese,chip89,par-simple:trics00,hent07:parallel-CSP,confid:cp09},
the problem is partitioned around the domains of the variables,
effectively partitioning its search space. The search for a solution
is then carried out in each of the sub-search spaces by one agent (or
worker), all agents working in parallel.

Constraint solving involves exploring large search spaces. To perform
search using several agents in parallel, the search effort must be
shared among them. This may happen either by having each agent do a
part of the work and coordinate with the other agents, in order to
reach the intended goal (which is the approach taken in solving
Distributed CSPs \cite{abt-yokoo98}), or the agents may be mostly
independent from each other, performing their (non-overlapping) part
of the work, hoping that one of them will find a quicker path to an
answer. While the first approach typically requires significant
inter-agent communication, not only for the search to progress but
also for termination detection, in the latter communication can be
limited to an initial dispatching of the agents and to an answer
collecting phase at the end of the procedure. In this case, however,
the initial work distribution may turn out to be quite unbalanced,
leaving some agents to bear most of the effort as others become idle
and their contribution is wasted.

This article reports on preliminary results of our experiments in
implementing a work-stealing scheme for overcoming the effect
described above. This is a two-level scheme: work stealing occurs
between co-located agents, but when distant agents are involved, some
cooperation is needed to redistribute the work still left.

The remainder of this paper is structured as follows: we start by
establishing some terminology in the next section. Then, in
Sections~\ref{sec:solver} and~\ref{sec:bench} we describe the
architecture of the implemented solver and report on some experimental
results obtained with it. Section~\ref{sec:related} discusses related
work and in Section~\ref{sec:conclusion} we conclude and put forward
possible continuation paths for this work.

\section{Constraint Solving}
\label{sec:csp}

A constraint satisfaction problem can be briefly defined as a set of
variables whose values, to be drawn from their domains, must satisfy a
set of relations.

\begin{definition}[CSP]
  A \emph{Constraint Satisfaction Problem} (CSP) over finite domains
  is a triple $P = (X,D,C)$, where
  \begin{itemize}
  \item $X = \{x_1, x_2, \ldots, x_n\}$ is an indexed set of
    \emph{variables};
  \item $D = \{D_1, D_2, \ldots, D_n\}$ is an indexed set of finite
    sets of values, with $D_i$ being the \emph{domain} of variable
    $x_i$, for every $i = 1,2,\ldots,n$; and
  \item $C = \{c_1, c_2, \ldots, c_m\}$ is a set of relations between
    variables, called the \emph{constraints}.
  \end{itemize}
\end{definition}

The \emph{search space} of a CSP consists of all the tuples from the
cross product of the domains, where each variable is assigned a value
from its domain. Solving a CSP amounts to finding some or all of those
tuples which satisfy all constraints of the problem.

\begin{definition}[Solution]
  A \emph{solution} to a CSP is an $n$-tuple $(v_1,v_2,\ldots,v_n) \in
  D_1 \times D_2 \times \ldots \times D_n$ such that all constraints
  are satisfied.
\end{definition}

In parallel constraint solving, the problem is divided into
subproblems. Solutions to these subproblems are also solutions to the
original problem.

\begin{definition}[Subproblem]
  A \emph{subproblem} of a CSP $P = (X,D,C)$ is a CSP $P' = (X,D',C)$
  such that $D' = \{D'_1, D'_2, \ldots, D'_n\}$ and $D'_i \subseteq
  D_i$, for every $i = 1,2,\ldots,n$.
\end{definition}

To guarantee completeness of the search, the search spaces of the
subproblems must cover the search space of the original problem. In
order to avoid redundant work, they must also be pairwise disjoint.

\begin{definition}[Partition]
  A set $\{P'_1, P'_2, \ldots, P'_k\}$ of subproblems of a CSP $P$,
  with $P'_i = (X,\{D'_{i1}, D'_{i2}, \ldots, D'_{in}\}, C)$, is a
  \emph{partition} of $P$ if
  \[
    \bigcup_{1 \le i \le k} D'_{i1} \times D'_{i2} \times \cdots
      \times D'_{in} = D_1 \times D_2 \times \cdots \times D_n
  \]
  and $(\forall\, i \ne j)\, D'_{i1} \times D'_{i2} \times \cdots
  \times D'_{in} \cap D'_{j1} \times D'_{j2} \times \cdots \times
  D'_{jn} = \emptyset\mbox{.}$
\end{definition}

A partition of a CSP may be dually regarded as a partition of its
search space, the search spaces of the subproblems being sub-search
spaces of the original problem. In this paper we will only deal with
search space partitions that correspond to some partition of a
problem.

\section{Solver Architecture}
\label{sec:solver}

\begin{figure}[h]
  \centering
  {
    \sffamily
    \begin{picture}(206,111)
      \put(36,105){Team 1}
      \put(0,0){\includegraphics[scale=1]{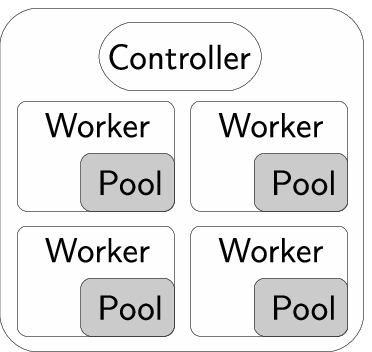}}
      \put(123.5,94){\scriptsize Team 2}
      \put(120,60){\includegraphics[scale=.3]{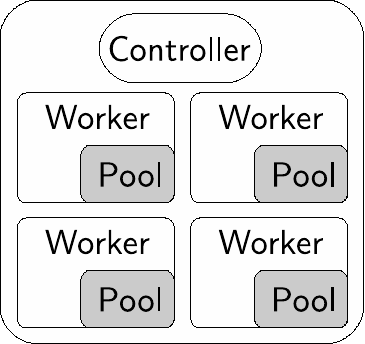}}
      \put(143.5,44){\scriptsize Team 3}
      \put(140,10){\includegraphics[scale=.3]{arch}}
      \put(178.5,79){\scriptsize Team 4}
      \put(175,45){\includegraphics[scale=.3]{arch}}
    \end{picture}
  }
  \caption{Solver architecture}
  \label{fig:arch}
\end{figure}

Our constraint solver consists of \emph{workers}, grouped together as
\emph{teams} (Figure~\ref{fig:arch}). The search for one or all
solutions is carried out by the workers, which implement a propagator
based constraint solving engine, following a domain consistency
oriented approach \cite{HCP:propagation}. Each active worker has a
\emph{pool} of \emph{idle} search spaces and a \emph{current} search
space, the one it is currently exploring. In each team there is a
\emph{controller}, which does not participate in the search, and one
of the controllers, the \emph{main controller}, also coordinates the
teams.

Structuring the workers this way serves two purposes: the first is
that a workers' sole task becomes searching, as all communication with
the environment required by the dynamic sharing of work among teams is
handled by the controller. The second objective is the sharing of
resources enabled by binding the workers in a team close together. If
all workers were on the same level, they would either have to divide
their attention between search and communication or there would have
to be one controller per worker, thereby increasing resource usage. On
the other hand, this structure matches naturally a two-level
partitioning of the search space and we obtain receiver-initiated
decentralised dynamic load balancing \cite{wilkinson+allen-2ed}.

At the outset of the search process, the problem to be solved is
partitioned and each team is entrusted with trying to solve one of the
resulting subproblems. The controller in each team then partitions the
local problem and hands each sub-search space over to a worker for
exploration.

On finishing exploring its assigned search space, a worker tries to
steal work from another worker \emph{within its team}. If
unsuccessful, it then notifies the team controller that it has become
idle. When all the workers in a team are idle, the controller asks the
other teams for more work.

\subsection{Partitioning Strategies}

The strategy used to partition the search space has a decisive impact
on the number of steps needed to get to a solution, hence on
performance.

Partitioning strategies may be designed either to lead to a balanced
distribution of the search work, like the \emph{even} strategy below
and the prime and greedy strategies from \cite{pas-silaghi01}, or to
produce some subproblems where the search is expected to be quick
(while others may be slow), such as \emph{eager} partitioning. In
principle, the former strategies will be more suited to situations
where all solutions are requested and the whole search space must be
visited, and the latter will lend themselves better to when one
solution is enough. In any case, the splitting of the problem will
introduce a breadth-first component into the usual depth-first
exploration of the search tree, which sometimes gives rise to
superlinear speedups.

In \emph{even partitioning}, domains are split so as to obtain
sub-search spaces of similar dimensions. If we want to split a problem
into $k$ subproblems, then the first variable with at least that many
values in its domain is chosen and its domain is split as evenly as
possible among the subproblems: if the domain of the chosen variable
has $d \ge k$ values, then it will have $\lfloor d / k \rfloor$ values
in the first $k - d \, \textbf{mod} \, k$ subproblems and $\lfloor d /
k \rfloor + 1$ values in the remaining $d \, \textbf{mod} \, k$
subproblems.

\emph{Eager partitioning} corresponds roughly to a partial
breadth-first expansion of the search tree and it will mostly produce
subproblems where at least one of the variables has had its domain
reduced to a single value. The splitting is performed according to the
algorithm depicted in Figure~\ref{algo:eager}, whose inputs are the
number of subproblems to create and a sequence of problems from which
to create them. Initially, this sequence only contains the original
problem.

\begin{figure}[ht]
  {
    \normalsize
    \begin{notation}
      If $P$ is a CSP and $D$ is a finite set, $PD_i$ stands for the
      CSP which is identical to $P$ except that the domain of the
      $i$\textsuperscript{th} variable is $D$.
    \end{notation}
  }
  {
    \sffamily
    \begin{tabbing}
      m \= m \= m \= m \= m \= m \= m \= \kill
      eager-split$(k, (P_1 \: P_2 \cdots P_q))$ \+ \\
      $(X, D, C) \leftarrow P_1$ \\
      $i \leftarrow \min \, \{ j \mid |D_j| > 1 \}$ \\
      $d \leftarrow |D_i|$ \\
      $\{v_1, v_2, \ldots, v_d\} \leftarrow D_i$ \\
      \IF\ $k \le d$ \THEN \+ \\
        $(P_1\{v_1\}_i \; P_1\{v_2\}_i \cdots P_1\{v_k, \ldots, v_d\}_i \; P_2 \cdots P_q)$ \- \\
      \ELSE \+ \\
        eager-split$(k-d+1, (P_2 \cdots P_q \; P_1\{v_1\}_i \; P_1\{v_2\}_i \cdots P_1\{v_d\}_i))$
    \end{tabbing}
  }
  \caption{Eager partitioning algorithm}
  \label{algo:eager}
\end{figure}

\begin{figure}[h]
  \centering
  \includegraphics[scale=1]{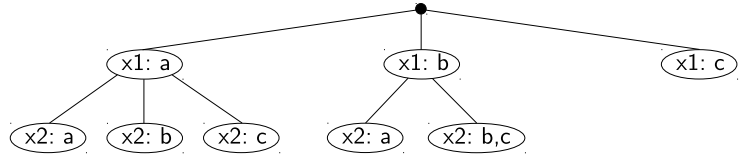}
  \caption{Eager partition into 6 sub-search spaces}
  \label{fig:eager}
\end{figure}

Figure~\ref{fig:eager} shows the result of applying eager partitioning
to split into six a problem where the domain of all variables is
$\{\textsf{a}, \mathsf{b}, \mathsf{c}\}$. (Only the variables whose
domains are affected by the splitting are shown.)

The partitioning of the CSP may affect the behaviour of the search,
even to the point of defeating the variable and value selection
heuristics which are usually appropriate to a given problem, as has
been noted in \cite[Section~6]{hent07:parallel-CSP}. This suggests
that the partitioning strategy, introducing another degree of freedom
in the search strategy, needs to be adapted to the problem being
solved and matched with the search heuristics used, and that no
overall `best' partitioning strategy exists. (Notice that, for the
present, problem specific heuristics do not inform problem
partitioning.)

As problem partitioning takes place at two points in the process ---
to distribute work to all the teams, and, initially within every team,
to assign work to each worker --- different splitting strategies can
be used, a more balanced one to allot similar amounts of work to the
individual teams, and another to focus the efforts of the agents. The
latter strategy could be finer grained than the former, the cost of
local work stealing being much lower than that of network supported
work sharing.

Additionally, in parallel search, different teams might split their
problems differently, allowing us to take advantage of one not yet
identified strategy being more effective than the others for the
problem at hand.

\subsection{Search}

The search unfolds as a worker further splits the search space it is
working on, keeping one part as its current search space and adding
the other to its pool of idle search spaces. If the current search
space is found to contain no solution, the worker draws a new search
space from the pool and starts exploring it, never backtracking. Upon
finding a solution, the worker communicates it to the team controller
which, in turn, forwards it to the main controller.

\begin{figure}[h]
  \centering
  \begin{picture}(230,141)(-24,49)
    \sffamily

    \put(117,156){
      \begin{tabular}[b]{l|c|}
        \cline{2-2}
        x1 & 1..4\\\cline{2-2}
        x2 & 1..4\\\cline{2-2}
        x3 & 1..4\\\cline{2-2}
      \end{tabular}
    }

    \put(50,163){\shortstack{past \\ search spaces}}

      \put(131.7,155.7){\line(-2,-1){26.75}}

        \put(72.75,109.25){
          \begin{tabular}[b]{l|c|}
            \cline{2-2}
            x1 &    1\\\cline{2-2}
            x2 & 1..4\\\cline{2-2}
            x3 & 1..4\\\cline{2-2}
          \end{tabular}
        }

          \multiput(87.5,109)(-16.5,-8.25){2}{\line(-2,-1){10}}

            \put(29,63){
              \begin{tabular}[b]{l|c|}
                \cline{2-2}
                x1 &    1\\\cline{2-2}
                x2 &    2\\\cline{2-2}
                x3 & 1..4\\\cline{2-2}
              \end{tabular}
            }

            \put(-23,73){\shortstack{current \\ search space}}

          \multiput(104.5,109)(16.5,-8.25){2}{\line(2,-1){10}}

            \put(116.5,63){
              \begin{tabular}[b]{l|c|}
                \cline{2-2}
                x1 &    1\\\cline{2-2}
                x2 & 3..4\\\cline{2-2}
                x3 & 1..4\\\cline{2-2}
              \end{tabular}
            }

      \put(149,155.7){\line(2,-1){26.75}}

        \put(161.25,109.25){
          \begin{tabular}[b]{l|c|}
            \cline{2-2}
            x1 & 2..4\\\cline{2-2}
            x2 & 1..4\\\cline{2-2}
            x3 & 1..4\\\cline{2-2}
          \end{tabular}
        }

    \put(160,100){\oval(90,100)}
    \put(175,60){Pool}
  \end{picture}
  \caption{Search spaces from a worker}
  \label{fig:search-spaces}
\end{figure}
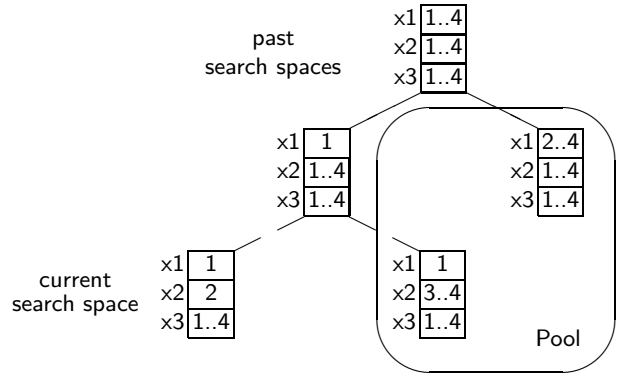

The state of a worker with two search spaces currently in the pool is
shown in Figure~\ref{fig:search-spaces}, where solid edges mean that
the child search spaces form a partition of the parent. Notice that
the subtree to the left of the current search space (corresponding to
the tuples where both $x_1$ and $x_2$ take value 1) has already been
explored and discarded, and is not displayed.

Figure~\ref{algo:worker} depicts the main driver algorithm for
workers. At each step of the search process, a worker starts by
looking within its current search space for a variable whose domain is
not a singleton (line~\ref{worker:sel}). If none is found, then the
search space contains a single tuple which constitutes a solution to
the problem, and which is returned by the worker
(line~\ref{worker:sol}). Otherwise, one of the variables with a
non-singleton domain is selected and the current search space is split
into two subspaces (line~\ref{worker:split}):
\begin{itemize}
\item In the first, which will become the worker's current search
  space, the selected variable is set to an individual value picked
  from its domain.
\item In the other, to be added to the pool of idle search spaces
  (line~\ref{worker:put}), that value is removed from the domain of
  the variable.
\end{itemize}
The domains of the other variables remain unchanged in both search
spaces.

Following the split, the new current search space goes through a
propagation phase (line~\ref{worker:prop}). If it succeeds, another
search step is performed. If the propagation fails, the worker tries
to fetch an idle search space from the pool to become the current
search space (line~\ref{worker:get}). If this is not possible the
worker fails (line~\ref{worker:fail}), otherwise the search resumes
with the retrieved search space undergoing a propagation phase, as the
domain of one of its variables shrunk just prior to it being stored in
the idle pool.

\begin{figure}[h]
  \sffamily

  \begin{algorithmic}[1]
    \PROC WORKER(search-space)
      \State current $\leftarrow$ search-space
      \While{var $\leftarrow$ select-variable(current)}     \label{worker:sel}
        \State (current, other) $\leftarrow$ split-search-space(var, current)
                                                            \label{worker:split}
        \State pool-put(other, var)                         \label{worker:put}
        \While{(current $\leftarrow$ revise(var, current)) = FAIL}
                                                            \label{worker:prop}
          \State (current, var) $\leftarrow$ pool-get()     \label{worker:get}
          \If{current = FAIL}
            \State \RETURN\ FAIL                            \label{worker:fail}
          \EndIf
        \EndWhile
      \EndWhile
      \State \Return SOLUTION(current)                      \label{worker:sol}
    \CORP
  \end{algorithmic}

  \rmfamily
  \caption{Worker main driver algorithm}
  \label{algo:worker}
\end{figure}

\subsection{Work Stealing}
\label{sec:steal}

When a worker tries to fetch a new search space from its pool and
finds it empty, it will attempt to obtain one from one of its
teammates. In order to minimise the impact on the performance of the
solver, this is achieved with as little cooperation from the holder of
the retrieved search space as possible. In fact, the idle worker will
effectively steal work from a teammate while the latter continues its
task, oblivious to what is being done to its work queue.

The intended discipline of a worker's pool is that of a \emph{deque}
(double-ended queue), as depicted in Figures~\ref{algo:pool-put-get}
and \ref{algo:steal}. While the owner works on one end of its pool
(lines~\ref{pool:put}, \ref{pool:rlast1}, and \ref{pool:rlast2}), a
worker whose pool is empty will remove an entry from the other end
(line~\ref{pool:rfirst}). This way, the only penalty a worker incurs
during normal processing is the cost of an extra check on the size of
its pool (line~\ref{pool:safe}). The protocol used to avoid
interference during pool accesses is similar to the one in
\cite{cilk-5}. Only when the number of entries in the pool is small,
will it be necessary to enforce mutual exclusion in the accesses to
the pool, and even then only when removing a search space. To reduce
contention, work stealing is only allowed from a pool when the number
of entries in it reaches a given threshold (line~\ref{pool:thresh}).

\begin{figure}[h!t]
  \centering
  {
    \sffamily
    \begin{algorithmic}[1]
      \PROC pool-put(search-space, variable)
        \State pool.append(search-space, variable)         \label{pool:put}
      \CORP

      \Statex

      \PROC pool-get()
        \If{pool.size = 0}
          \State \Return steal-work()
        \ElsIf{pool.size $<$ SAFE-SIZE}                    \label{pool:safe}
          \State lock(pool)
          \State ss $\leftarrow$ pool.remove-last()        \label{pool:rlast1}
          \State unlock(pool)
          \State \Return ss
        \Else
          \State \Return pool.remove-last()                \label{pool:rlast2}
        \EndIf
      \CORP
      \algstore{pool:break}
    \end{algorithmic}
  }
  \caption{Pool insertion and removal}
  \label{algo:pool-put-get}
\end{figure}

\begin{figure}[h!t]
  \centering
  {
    \sffamily
    \begin{algorithmic}[1]
      \algrestore{pool:break}
      \PROC steal-work()
        \State lock(stealing)
        \State v $\leftarrow$ worker-with-biggest-pool()
        \State lock(v.pool)
        \If{v.pool.size $<$ THRESHOLD}                     \label{pool:thresh}
          \State ss $\leftarrow$ FAIL
        \Else
          \State ss $\leftarrow$ v.pool.remove-first()     \label{pool:rfirst}
        \EndIf
        \State unlock(v.pool)
        \State unlock(stealing)
        \State \Return ss
      \CORP
    \end{algorithmic}
  }
  \caption{Work stealing algorithm}
  \label{algo:steal}
\end{figure}

Stolen work corresponds to locations nearer the root of a worker's
search tree. The search within the worker's search space proceeds
according to the heuristics deemed adequate to the problem until it
either finds a solution or the work is exhausted. Upon stealing work
from a peer, a worker picks up the search at a point that the worker
it was stolen from would eventually reach, thus subverting the
problem's search strategy and introducing in it a measure of
randomness. This may be either beneficial or detrimental, depending on
the specific problem.

In the event of an idle worker failing to obtain work within its team,
it notifies the team controller and waits, either to be later restarted or
to be terminated. When all the agents in a team have become idle, the
team controller broadcasts a request for more work to the other teams.

Inter-team work stealing follows along a simple plan: initially, one
of the team controllers is given the role of fulfilling requests for
work. Upon receiving one, and using the same protocol used by the
workers, it tries to steal a search space from the local pool to be
forwarded to the requester, which splits it among its workers and
becomes the new work supplier. If the designated work supplier is
unable to spare a search space, the remaining teams are polled for
work, as done in \cite{par-simple:trics00}. When no team is able to
supply additional work, the idle team notifies the main controller and
terminates.

\subsection{Implementation Notes}

One of the main goals behind this work was to build a constraint
solver which could take advantage of the advances in parallel
architectures and in clustering network technology. To better be able
to handle the challenges inherent to multiprocessing, namely memory
management and caching issues, C was our choice for the implementation
language, as it allows for very fine-grained control.

A key idea behind the implementation is that of \emph{store}. A store
describes the domains of the variables of the problem and represents a
sub-search space of the initial problem. Stores constitute the state
of a worker and are meant to be self-contained and dense. The search
spaces in the idle pools of each worker are represented by stores,
which may be copied between workers and transmitted between teams,
thus allowing the redistribution of the search.

Teams are autonomous entities and each team corresponds to a distinct
process, usually residing on a dedicated machine. As communication,
particularly over a network, may have an adverse impact on system
performance, care has been taken to minimise the number of inter-team
messages needed. Teams are coordinated by way of an IPC library.

A team comprises active components which are the workers and the
controller. The controller is, most of the time, waiting for a worker
or another team controller to communicate with it, not disturbing the
search process and allowing workers to be mapped to processors.
Workers are mostly independent from each other, except where work
stealing is concerned, as explained in Section~\ref{sec:steal}. A
worker, to be able to steal work from another one without active
cooperation from the latter, must be able to access all the team
pools. To make this possible, pools are located in shared memory and
workers, as well as the controller, are implemented as lightweight
processes (threads).

\section{Experimental Results}
\label{sec:bench}

In this Section, we present some performance results obtained with our
solver on three classic benchmark problems, namely the non attacking
queens problem, the Golomb ruler problem \cite[problem 006]{csplib},
and the Langford number problem \cite[problem 024]{csplib}.
Measurements were made of the time taken to count all solutions for
the three problems and for generating the first solution in the latter
problem.

These measurements were made on a cluster of Q6600 Intel Core2 Quad
CPUs, clocked at 2.4GHz, with 2--4GB RAM, running Linux, and the code
was compiled with GCC 4.1.1 with the `-O3' flag. The times presented
are the average of the middle 10 times from 12 runs of each program.
When computing the relative performance with respect to the sequential
case, we subtracted the overhead associated with starting up and
terminating the solver, which reached a maximum of 0.2 seconds in the
6 teams configuration. Unless otherwise indicated, teams are composed
of 4 workers, mirroring the number of CPUs in the shared-memory
multiprocessor systems. For interprocess communication, the Open MPI
MPI-2 implementation \cite{open-mpi} was used.

Absolute performance has not, so far, been the top priority goal of
this work. Nevertheless the sequential (1 team with 1 worker) version
of our solver already displays interesting times for solving these
problems, as attested by Table~\ref{tab:times}, where they are
compared with those of Gecode \cite{gecode}, although there clearly
remains some work to be done in that regard.

\begin{table}[h]
  \centering
  \caption{Times comparison with Gecode (seconds)}
  \label{tab:times}
  \begin{tabular}{@{}|@{\,}c@{\,}|*{9}{r@{\,}|}@{}}
    \cline{2-10} \multicolumn{1}{@{}c|}{}
               & Golomb & \multicolumn{3}{c|}{Queens} & \multicolumn{5}{c|@{}}{Langford}     \\
    \multicolumn{1}{@{}c|}{}
               &     10 &    14 &     15 &     16     &  2 11 &  2 12 &  2 28 & 2 31 & 3 18  \\ \hline
    Our solver &   3.53 & 13.89 &  86.05 & 580.36     &  1.07 &  8.00 & 67.56 & 1.26 & 2.44  \\ \hline
    Gecode     &   0.85 & 17.21 & 102.18 & 646.43     & 36.40 & 25.01 &  0.03 & 0.02 & 0.42  \\ \hline
    \multicolumn{1}{c|}{}
               & \multicolumn{6}{c|}{all solutions} & \multicolumn{3}{c|@{}}{first solution} \\ \cline{2-10}
  \end{tabular}
\end{table}

The current implementation still suffers from some limitations which
restrict the range of problems we are able to run. One is the internal
representation of the domains, which only allows values between 0 and
63. Another is that we do not deal with optimisation constraints.
These are required by the typical formulation of the Golomb ruler
problem where, to make up for their absence, we bound the domains of
the variables from above by the known minimum value of the last mark
of the ruler.

In the remainder of this section, we look at the results obtained with
several configurations of the solver and analyse them with respect to
the speedups induced by the parallelisation of the search, using the
two partitioning strategies. The use of the two strategies helps to
highlight the effect of work stealing.

To study the effects of the parallelisation of the solving procedure
on the Golomb ruler problem with 10 marks, we measured the amount of
work associated with each value from the domain of the variable $m_1$
corresponding to the first positive mark of the ruler. This is the
variable around which the problem is partitioned.

Table~\ref{tab:golomb:work} shows the weight of exploring all the
tuples where variable $m_1$ takes one value from its domain within the
effort of exploring the whole search space. For instance, the time
taken to explore the subtree where $m_1$ has value 3 is 17.1\% of the
time needed to explore the whole tree, and the times for the first
three values of the domain together correspond to 66.5\% of the total
time. All values between 10 and 55, which is the minimum length of a
10-mark Golomb ruler, together account for about 1.4\% of the total
work effort. These values also reflect the amount of pruning that
takes place during the search.

\begin{table}[ht]
  \centering
  \caption{Work distribution in the Golomb ruler problem (10 marks, all
    solutions), with respect to the value of the first positive mark}
  \label{tab:golomb:work}
  \begin{tabular}{c|*{10}{r|}}
    %
    %
    $m_1$ & 1    & 2    & 3    & 4    & 5    & 6    & 7    & 8    & 9    & 10\,\ldots 55 \\ \hline
    \%    & 25.5 & 23.9 & 17.1 & 12.2 &  8.1 &  5.1 &  3.3 &  2.2 &  1.2 &   1.4         \\ \hline
          & 25.5 & 49.3 & 66.5 & 78.7 & 86.8 & 91.9 & 95.2 & 97.4 & 98.6 & 100.0         \\ \cline{2-11}
  \end{tabular}
\end{table}

\begin{figure}[ht]
  \centering
  \hspace{-7pt}%
  \includegraphics[scale=.775]{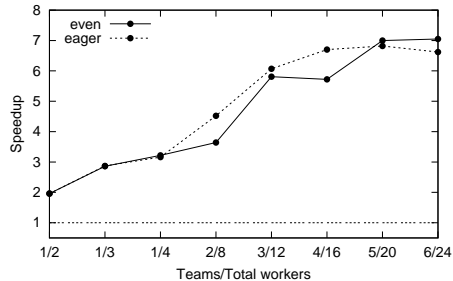}
  \caption{Speedups for the Golomb ruler (10 marks, all solutions)}
  \label{gra:golomb:process:speedups}
\end{figure}

From Figure~\ref{gra:golomb:process:speedups}, we can see that the
speedup increase is roughly linear up to 3 workers and improves up to
4 teams. Results are similar with both partitioning strategies, which
speaks for the effectiveness of the work stealing implementation. For
more than 4 teams, when total running times are around 0.7 seconds,
the overhead of communication takes over.

For up to 6 teams, even partitioning will make the domain of $m_1$ in
the first team correspond to at least 98.6\% of the work. Without work
sharing this would mean that the speedup would be bounded by 4. Work
sharing allows it to reach higher values, even with the simple scheme
employed. Given the structure of the problem, eager partitioning
divides the work more evenly among the subproblems and with 6 teams
the work distribution will be approximately 25--24--17--12--8--14\%.
The fact that the speedups are similar in both cases is due to work
sharing evening out these differences. However, in the latter
situation speedups of around 8 should be possible, even without
inter-team work sharing.

\begin{figure}[ht]
  \centering
  \hspace{-7pt}%
  \includegraphics[scale=.775]{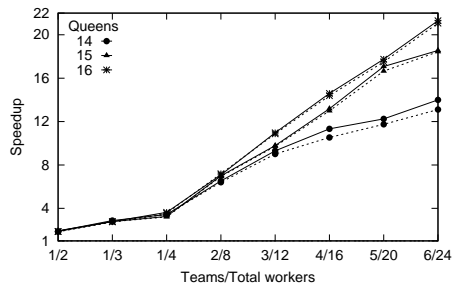}
  \caption{Speedups for the non attacking queens (all solutions)%
    \protect\footnotemark}
  \label{gra:queens:process:speedups}
\end{figure}

In the non attacking queens problem, the first observation that can be
made in relation to the speedups obtained, depicted in
Figure~\ref{gra:queens:process:speedups}, is that they are fairly
insensitive to the partitioning strategy used, which is a trend in the
results referred in this section. Given that in this problem the work
is very evenly distributed among the possible values from the domains
of the variables, this result is only possible due to effective work
sharing.

The profile of the speedups evolution with the addition of more teams
is illustrating in this case. While it is quasi-linear for the 16
queens problem, showing good scalability of the approach, the smaller
problem starts suffering from the weight of the implementation early
on. Total running times for the three problems in the 6 team setting
are around 1.2, 4.8, and 27.5 seconds, for 14, 15, and 16 queens,
respectively.

\begin{figure}[ht]
  \centering
  \subfloat[First solution]{%
    \hspace{-7pt}%
    \includegraphics[scale=.775]{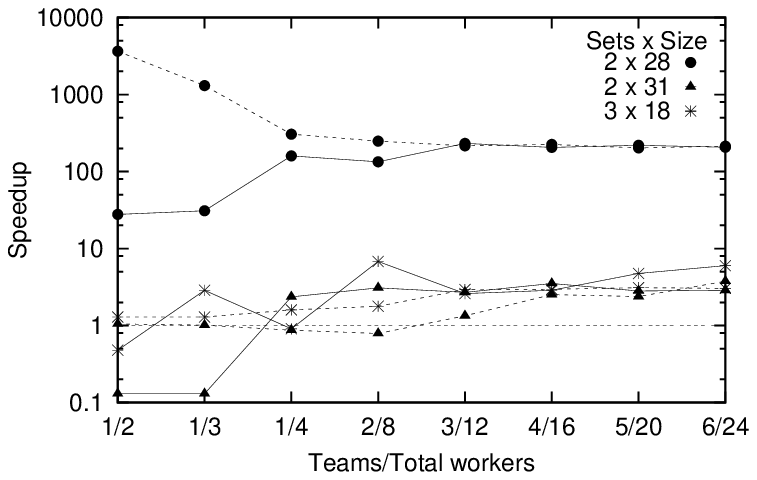}%
    \label{gra:langf-one:process:speedups}%
  }
  \subfloat[All solutions]{%
    \hspace{-7pt}%
    \includegraphics[scale=.775]{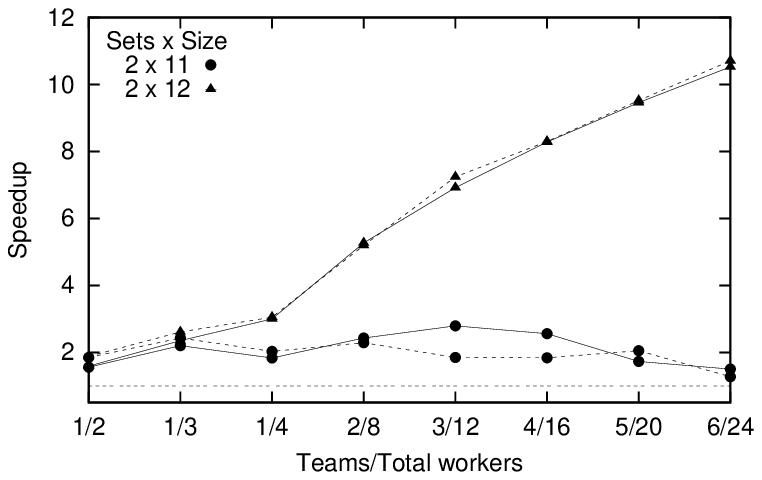}%
    \label{gra:langf-all:process:speedups}
  }
  \caption{Speedups for the Langford number problem}
\end{figure}

The Langford number problem, for which we measured both the speedups
for counting all solutions and for obtaining the first solution, is an
example of a case where domain partitioning interacts badly with the
heuristics usually used for guiding the search, as dividing a domain
gives rise to more work than that needed to solve the original
problem. This is apparent in
Figure~\ref{gra:langf-one:process:speedups}, which represents the
results observed in finding the first solution and where some
instances of the problem displayed a marked slowdown when partitioning
the domain of the first variable in two or three similarly size parts.
On the other hand, speedups of more than 3000 were also obtained in
one case.

\footnotetext{In these graphs, solid and dashed lines correspond,
  respectively, to even and eager partitioning.}

Counting all solutions of the Langford problem
(Figure~\ref{gra:langf-all:process:speedups}) exhibits a profile
common to the previous problems, where at some point the
implementation starts overwhelming the potential improvements due to
the parallelisation. This effect requires further study to identify
and solve its causes.

\section{Related Work}
\label{sec:related}

Our aim is to find how best to take advantage of the available
parallel architectures for constraint solving, taking into account
factors such as the search and problem partitioning strategies, as
well as communication and memory access patterns which may help in
minimising the overhead introduced by the competition for resources
involved in a parallel or distributed system and by the coordination
of the system components.

Recent years have seen an increase in the interest in parallel
solving, as parallel architectures become more common. An early
language sporting parallel constraint solving was the CHIP parallel
constraint logic programming language \cite{chip89}. It was
implemented on top of the logic programming system PEPSys, whose
or-parallel resolution infrastructure was adapted to handle the domain
operations needed in parallel constraint solving.

More recent works rely on features of an underlying framework for
programming parallel search. The concurrent Oz language provides the
basis for the implementation described in \cite{par-simple:trics00},
where search is encapsulated into computation spaces and a distributed
implementation allows the distribution of workers. Work sharing is
coordinated by a manager, which receives requests for work from the
workers and then tries to find one willing to share the work it has
left. Search strategies are user programmed and the work sharing
strategy is implemented by the workers.

A similar approach is taken in \cite{hent07:parallel-CSP,%
  hent08:parallel-CSP} which show how to program parallel search
controllers in \textsc{Comet}. There, the pool is an active object
which is queried by the idle workers. In case the pool is empty, it
asks another worker to generate yet unexplored sub-search spaces,
gives one away and stores the rest. It is not explained, however, how
the worker which supplies work is chosen.

A focus of research has been on the strategies for splitting the work
between workers. These strategies may be driven by the problem
structure, such as the size of the domains \cite{pas-silaghi01}, or by
the past behaviour of the solver, be it related with properties of the
solving process, such as the number of variables already instantiated
\cite{load-bal:rolf}, or with the progress of the search, in what it
affects the prospects of finding a solution in the current subtree
\cite{concdb-ai06}
or in the subtrees left to explore~\cite{confid:cp09}.

\section{Conclusions and Future Work}
\label{sec:conclusion}

Parallelisation seems to be a natural way of improving the performance
of CSP solving, and the results presented in this paper confirm the
gains it may produce. However, as the performance of sequential search
for different problems is highly dependent on the heuristics used, it
remains a challenge to identify the partitioning strategies which will
be more appropriate to the parallel solving of each problem. While it
may be tempting to adapt the sequential search heuristics to problem
splitting, the granularity they induce on the sub-search spaces may be
too fine and lead to too great a communication overhead in distributed
settings. So, there is a tradeoff to be struck between how closely
partitioning follows the `optimal' search order for a given problem
and the impact it has on the operation of the system.

In \cite{confid:cp09}, a scheme is presented which uses the search
heuristics to guide problem splitting, dampened by a degree of
confidence to distribute the workers across the search tree while
maintaining some bias towards the nodes favoured by the heuristic. It
shows good performance on multi-core hardware, and while it has the
drawback of working on a global view of the search process, it seems
to point in a promising direction of research, namely using the work
done as a guide to future search space splitting.

In spite of the results obtained so far, there should be additional
gains with a more sophisticated work sharing protocol. Several
possibilities should be studied, including having a different work
stealing policy for inter-team sharing, where candidate search spaces
undergo a deeper examination to try to determine whether the cost of
their sending is offset by the work saved locally.

Planned developments of this work, besides tackling the defects and
limitations identified in this text, comprise the inclusion of
optimisation constraints and the improvement of the scalability of the
implementation in two key aspects: the initial work distribution and
the sharing of work between teams, which could both profit from
organising the teams in multi-level neighbourhoods.

We plan on experimenting with different underlying models and
libraries for thread management and inter-process communication,
namely to venture beyond the present implementation which relies on
Posix threads and MPI.

\section*{Acknowledgements}

The authors wish to acknowledge the FCT/Pessoa grant `CONTEMP ---
CONTraintes Ex\'ecut\'ees en MultiProcesseurs' and the members of the
partner INRIA/Bordeaux RUNTIME team, namely Olivier Aumage, J\'er\^ome
Clet-Ortega, and C\'edric Augonnet, for their cooperation and helpful
suggestions. Thanks are due to Miguel Avillez at Universidade de
\'Evora for the valued offer of computational support. The authors
would also like to thank the anonymous referees for extended comments
and for suggesting ways to improve this work.

\bibliography{strings,restricoes,concorrencia,arquitectura}
\bibliographystyle{splncs03}

\end{document}